\documentclass[12pt]{iopart}
\usepackage{iopams,graphicx}

\newcommand{\nc}{\newcommand}

\nc{\be}[1]{\begin{equation}\mbox{$\label{#1}$}}
\nc{\bea}[1]{\begin{eqnarray} \mbox{$\label{#1}$}}
\nc{\Section}[2]{\section{#2}\label{#1}}
\nc{\Bibitem}[1]{\bibitem{#1}} \nc{\Label}[1]{\label{#1}}

\nc{\eea}{\end{eqnarray}} \nc{\ee}{\end{equation}}

\nc{\bdm}{\begin{displaymath}} \nc{\edm}{\end{displaymath}}
\nc{\dpsty}{\displaystyle} \nc{\bc}{\begin{center}}
\nc{\ec}{\end{center}} \nc{\ba}{\begin{array}} \nc{\ea}{\end{array}}
\nc{\bab}{\begin{abstract}} \nc{\eab}{\end{abstract}}
\nc{\btab}{\begin{tabular}} \nc{\etab}{\end{tabular}}
\nc{\bit}{\begin{itemize}} \nc{\eit}{\end{itemize}}
\nc{\ben}{\begin{enumerate}} \nc{\een}{\end{enumerate}}
\nc{\bfig}{\begin{figure}} \nc{\efig}{\end{figure}}

\nc{\arreq}{&\!=\!&} \nc{\arrmi}{&\!-\!&} \nc{\arrpl}{&\!+\!&}
\nc{\arrap}{&\!\!\!\approx\!\!\!&} \nc{\non}{\nonumber}
\nc{\align}{\!\!\!\!\!\!\!\!&&}

\def\lsim{\; \raise0.3ex\hbox{$<$\kern-0.75em
     \raise-1.1ex\hbox{$\sim$}}\; }
\def\gsim{\; \raise0.3ex\hbox{$>$\kern-0.75em
     \raise-1.1ex\hbox{$\sim$}}\; }

\nc{\DOT}{\hspace{-0.08in}{\bf .}\hspace{0.1in}} \nc{\Laada}{\hbox
{$\sqcap$ \kern -1em $\sqcup$}}
\nc\loota{{\scriptstyle\sqcap\kern-0.55em\hbox{$\scriptstyle\sqcup$}}}
\nc\Loota{{\sqcap\kern-0.65em\hbox{$\sqcup$}}} \nc\laada{\Loota}
\nc{\qed}{\hskip 3em \hbox{\BOX} \vskip 2ex}

\nc{\real}{{\rm I \! R}} \nc{\Z}{{\sf Z \!\!\! Z}}
\nc{\complex}{{\rm C\!\!\! {\sf I}\,\,}}
\def\bigid{\leavevmode\hbox{\small1\kern-3.8pt\normalsize1}}
\def\id{\leavevmode\hbox{\small1\kern-3.3pt\normalsize1}}
\nc{\slask}{\!\!\!/} \nc{\bis}{{\prime\prime}} \nc{\pa}{\partial}
\nc{\na}{\nabla} \nc{\ra}{\rangle} \nc{\la}{\langle}
\nc{\goto}{\rightarrow} \nc{\swap}{\leftrightarrow}

\nc{\EE}[1]{ \mbox{$\cdot10^{#1}$} } \nc{\abs}[1]{\left|#1\right|}
\nc{\at}[2]{\left.#1\right|_{#2}} \nc{\norm}[1]{\|#1\|}
\nc{\abscut}[2]{\Abs{#1}_{\scriptscriptstyle#2}}
\nc{\vek}[1]{{\rm\bf #1}} \nc{\integral}[2]{\int\limits_{#1}^{#2}}
\nc{\inv}[1]{\frac{1}{#1}} \nc{\dd}[2]{{{\partial #1}\over{\partial
#2}}} \nc{\ddd}[2]{{{{\partial}^2 #1}\over{\partial {#2}^2}}}
\nc{\dddd}[3]{{{{\partial}^2 #1}\over{\partial #2 \partial #3}}}
\nc{\dder}[2]{{{d #1}\over{d #2}}} \nc{\ddder}[2]{{{d^2 #1}\over{d
{#2}^2}}} \nc{\dddder}[3]{{d^2 #1}\over
   {d #2 d #3}}
\nc{\dx}[1]{d\,^{#1}x} \nc{\dy}[1]{d\,^{#1}y} \nc{\dz}[1]{d\,^{#1}z}
\nc{\dl}[1]{\frac{d\,^{#1}l}{(2\pi)^{#1}}}
\nc{\dk}[1]{\frac{d\,^{#1}k}{(2\pi)^{#1}}}
\nc{\dq}[1]{\frac{d\,^{#1}q}{(2\pi)^{#1}}}

\nc{\bfT}{{\bf T }}

\nc{\cA}{{\cal A}} \nc{\cB}{{\cal B}} \nc{\cD}{{\cal D}}
\nc{\cE}{{\cal E}} \nc{\cG}{{\cal G}} \nc{\cH}{{\cal H}}
\nc{\cL}{{\cal L}} \nc{\cO}{{\cal O}} \nc{\cT}{{\cal T}}
\nc{\cN}{{\cal N}} \nc{\cR}{{\cal R}}
\nc{\rvac}[1]{|{\cal O}#1\rangle} \nc{\lvac}[1]{\langle{\cal O}#1|}
\nc{\rvacb}[1]{|{\cal O}_\beta #1\rangle}
\nc{\lvacb}[1]{\langle{\cal O}_\beta #1 |} \nc{\bb}{\bar{\beta}}
\nc{\bt}{\tilde{\beta}} \nc{\ctH}{\tilde{\cal H}}
\nc{\chH}{\hat{\cal H}}
\nc{\1}{\aa} \nc{\2}{\"{a}} \nc{\3}{\"{o}} \nc{\4}{\AA}
\nc{\5}{\"{A}} \nc{\6}{\"{O}}
\nc{\al}{\alpha} \nc{\g}{\gamma} \nc{\Del}{\Delta}
\nc{\eps}{\epsilon} \nc{\lam}{\lambda} \nc{\Om}{\Omega}
\nc{\ve}{\varepsilon} \nc{\mn}{{\mu\nu}} \nc{\vp}{\varphi}

\nc{\rf}[1]{(\ref{#1})} \nc{\nn}{\nonumber \\*} \nc{\bfB}{\bf{B}}
\nc{\bfv}{\bf{v}} \nc{\bfx}{\bf{x}} \nc{\bfy}{\bf{y}}
\nc{\vx}{\vec{x}} \nc{\vy}{\vec{y}} \nc{\oB}{\overline{B}}
\nc{\oI}{\overline{I}} \nc{\oR}{\overline{R}} \nc{\rar}{\rightarrow}
\nc{\ti}{\times} \nc{\slsh}{\hskip-5pt/} \nc{\sm}{Standard~Model~}
\nc{\MP}{M_{\rm Pl}} \nc{\mpl}{M_{\rm Pl}} \nc{\tp}{t_{\rm Pl}}

\nc{\pmin}{p_{\rm min}} \nc{\pmax}{p_{\rm max}} \nc{\fo}{f_0}
\nc{\foi}{f_{0,i}\,} \nc{\fop}{f_0^P} \nc{\fou}{f_0^U}

\nc{\eff}{{\rm eff}} \nc{\MT}{M_{\rm T}} \nc{\ML}{M_{\rm L}}
\nc{\kk}{\vek{k}} \nc{\pp}{{\rm p}} \nc{\half}{{1\over 2}}
\nc{\w}{\omega} \nc{\uhat}{\hat{U}_\w}

\nc{\ie}{{\it i.e. }} \nc{\eg}{{\it e.g. }} \nc{\trh}{T_{\rm RH}}
\nc{\ad}{{a'\over a}} \nc{\bd}{{b'\over b}} \nc{\Rd}{{R'\over R}}
\nc{\diag}{{\textrm{diag}}} \nc{\mato}[1]{\tilde{#1}}
\nc{\sinn}{\textrm{sinn}}
\nc{\sech}{\textrm{sech}}
\nc{\I}{\textrm{I}} \nc{\II}{\textrm{II}} \nc{\III}{\textrm{III}}
\nc{\vev}[1]{\langle #1 \rangle} \nc{\hyp}{\,\; F_{1{\hskip
-16pt}2}{\hskip 11pt}} \nc{\brhom}{\overline{\rho}_M}
\nc{\rhob}{\overline{\rho}} \nc{\Pb}{\overline{P}}
\nc{\bH}{\overline{H}} \nc{\ep}{{1+4\eps}}

\def\smiley{\hbox{\large$\bigcirc$\hspace{-.80em}%
\raise.2ex\hbox{$\cdot\cdot$}\kern-.61em    
\lower.2ex\hbox{\scriptsize$\smile$}}\ }

\def\frowney{\hbox{\large$\bigcirc$\hspace{-.80em}%
\raise.2ex\hbox{$\cdot\cdot$}\kern-.635em
\lower.2ex\hbox{\scriptsize$\frown$}}\ }

\begin{document}

{\title[Bayesian Analysis and Constraints on Kinematic Models from
Union SNIa]{Bayesian Analysis and Constraints on Kinematic Models 
from Union SNIa}}

\author{A.C.C. Guimar\~aes, J.V. Cunha and J.A.S. Lima}

\address{Departamento de Astronomia,  Universidade de S\~ao Paulo,\\ 
Rua do Mat\~ao 1226, CEP 05508-090 S\~ao Paulo SP, Brazil}

\eads{\mailto{aguimaraes@astro.iag.usp.br},
\mailto{cunhajv@astro.iag.usp.br}, \mailto{limajas@astro.iag.usp.br}}

\begin{abstract}
The kinematic expansion history of the universe is investigated by using 
the 307 supernovae type Ia from the Union Compilation set.  
Three simple model parameterizations for the deceleration parameter 
(constant, linear and abrupt transition) and two different models 
that are explicitly parametrized by the cosmic jerk parameter 
(constant and variable) are considered.  
Likelihood and Bayesian analyses are employed to find 
best fit parameters and compare models among themselves and with the
flat $\Lambda$CDM model.
Analytical expressions and estimates for the deceleration and cosmic jerk 
parameters today ($q_0$ and $j_0$) and for the 
transition redshift ($z_t$) between a past phase of cosmic deceleration 
to a current phase of acceleration are given.
All models characterize an accelerated expansion for the
universe today and largely indicate that it was decelerating in
the past, having a transition redshift around 0.5.
The cosmic jerk is not strongly constrained by the present 
supernovae data.
For the most realistic kinematic models the $1\sigma$ confidence limits
imply the following ranges of values: $q_0\in[-0.96,-0.46]$, 
$j_0\in[-3.2,-0.3]$ and $z_t\in[0.36,0.84]$,
which are compatible with the $\Lambda$CDM predictions, $q_0=-0.57\pm0.04$, 
$j_0=-1$ and $z_t=0.71\pm0.08$.
We find that even very simple kinematic models are equally good to 
describe the data compared to the concordance $\Lambda$CDM model, 
and that the current observations are not powerful enough to discriminate 
among all of them.  
\end{abstract}
\noindent{\it Keywords\/}: supernova type Ia - standard candles, dark energy experiments, dark energy theory

\section{Introduction}

The extension of the Hubble diagram to larger distances by using
observations from supernovae type Ia (SNIa) as standard candles,
allowed the history of cosmic expansion to be probed with deeper
detail. Independent measurements by various groups indicated that
the current expansion is in fact speeding up and not slowing down, 
as believed for many decades 
\cite{1998AJ....116.1009R,perlmutter,riess04,snls}. 
In other words, in virtue of some unknown mechanism, the expansion 
of the Universe underwent a ``dynamic phase transition" whose 
main effect is to change the sign of the universal deceleration 
parameter $q(z)$.

The physical explanation for such a transition is one of the greatest
challenges for cosmology today. Inside the General Relativity
paradigm, it requires the presence of a cosmological constant in the
cosmic equations, or to postulate the existence of an exotic fluid
with negative pressure (in addition to dark matter),  usually called
dark energy \cite{rev1,rev2,rev3,rev4,rev5}, or even a
gravitationally-induced cold dark matter creation
\cite{lima96,lima08,gary08}. Another possibility is to change the
theory describing the gravitational interaction as happens, for
instance, in the framework of the so-called F(R) modified gravity
theories \cite{FR,FR1,FR2,FR3,FR4}. 
In both cases, the space parameter associated with the cosmic expansion 
is too degenerate, and, as such, it is not possible, based on the 
current data, to decide which mechanism or dark energy component is 
operating in the cosmic dynamics \cite{Komat08}.

Another very distinct firsthand approach to access the history of
the cosmic expansion without the use of quantities coming from the 
dynamic description has also been proposed in the literature 
\cite{TurRie02}. 
This route is very interesting because it depends neither on the 
validity of any particular metric theory of gravity  nor on the 
matter-energy content of the observed Universe. 
It is closely related to the weaker assumption that space-time is
homogeneous and isotropic, so that the FRW metric is still valid, as
are the kinematic equations for redshift/scale factor. 
Some call it cosmography \cite{Visser04,shapiro} or 
cosmokinetics \cite{blandford05}, others use the term
Friedmannless cosmology \cite{EM1,EM}, but, in what follows,
we refer to it simply as a kinematic approach since it holds true
regardless of the underlying cosmic dynamics
\cite{riess04,Virey05,rapp07,daly08,CL08,Cunha09}.

Few years ago, Elgar{\o}y \& Multam{\"a}ki \cite{EM} investigated
constraints on some kinematic models by employing a Bayesian
marginal likelihood analysis based on the Gold  Supernova
sample data of Riess et al. \cite{riess04} and the Supernova Legacy
Survey (SNLS) of Astier et al. \cite{snls}. In their analysis of the
flat case, three different parameterizations for an accelerating $q(z)$
model were examined, namely: constant, linear,  and abrupt
transition, respectively, $M_0$, $M_1$ and $M_2$ in their
nomenclature. It was also argued that any expansion of the jerk
parameter (the third order contribution in the expansion for
kinematic luminosity distance in terms of the redshift z) could be
seen as requiring more parameters, or cosmic fluids in the dynamic 
approach, than expanding
the deceleration parameter. In addition, it was also observed that
the flat $\Lambda$CDM and the Einstein-de Sitter model have constant jerk
parameter $j_0=-1$, and, therefore, it cannot be used for
discriminating such cosmologies. Accordingly,  their
analysis was restricted to the deceleration parameter.

In this work we go one step further by examining the case for
the jerk parameter (constant and variable). One basic reason is 
that the bi-dimensional space parameter ($q_0,j_0$) can naturally 
discriminate the flat $\Lambda$CDM and Einstein-de Sitter models 
because they have different predictions for $q_0$. 
Actually, it is not necessary to expand $j(z)$, thereby introducing 
many parameters in order to have a larger class of models, since 
this happens even for constant jerk.
Probably, and more importantly, it is not clear a priori if the Bayesian
evidence prefers the cosmic concordance model when a more general
constant jerk parameter is considered. Potentially, a constant jerk
parameter provide us with the simplest approach to search for
departures from the cosmic concordance model. For completeness, in
our analysis we also consider all the flat models discussed by
Elgar{\o}y \& Multam{\"a}ki \cite{EM}, however, we also examine the
predictions of two different approaches including  constant and variable jerk
parameters. In addition, and differently from previous works, the present
Bayesian analysis is based on the larger Union Compilation SNIa
data recently published by Kowalski et al. \cite{kowalski2008}.

\section{Kinematic Models}

Assuming that the universe is homogeneous and isotropic above some
scale, the Friedmann-Robertson-Walker (FRW) metric  provides a good
description of the geometry of the universe ($c=1$)
\be{frw} ds^2=dt^2-a^2(t)\left[\frac{dr^2}{1-k
r^2}+r^2d\Omega\right],
\ee
where $k$ is the space curvature that we
will assume hereafter to be null, since the current emerging consensus
is that the universe is flat or very close to flat \cite{Komat08}.
The function $a(t)$ is the scale factor,
which contains the complete history of the cosmic expansion and can
be parameterized by the redshift, $a=(1+z)^{-1}$ ($a_0=1$).

The rate of expansion and the acceleration are represented by the Hubble
and deceleration parameters, respectively 
\be{hubble} H\equiv \frac{\dot{a}}a \,, \ee 
and
\be{dec} q\equiv -\frac{1}{H^2}\frac{\ddot{a}}{a}= \frac 12
(1+z)\frac{[H(z)^2]'}{H(z)^2}-1\,. \ee 
Similarly, the jerk parameter is defined as 
\be{jerk} j\equiv -\frac{1}{H^3}
\frac{\dot{\ddot{a}}}{a}=-\left[\frac 12 (1+z)^2
\frac{[H(z)^2]''}{H(z)^2}-(1+z)\frac{[H(z)^2]'}{H(z)^2}+1\right].
\ee 
The basic aim here is to examine some simple kinematic models
for the cosmic expansion based on specific parameterizations for
$q(z)$ in (\ref{dec}) and a constant jerk parameter.

The first and simplest model, $M_0$, is given by a constant
deceleration parameter, $q(z)=q_0$. 
The second model, $M_1$, is a linear expansion of the deceleration 
parameter $q(z)=q_0+q_1z$ (first used by \cite{riess04}).
Model $M_2$ (introduced by \cite{TurRie02}) depicts two phases of 
constant deceleration parameter, separated by an abrupt transition 
redshift, $q(z)=q_0$ for $z\leq z_t$ and $q(z)=q_1$ for $z> z_t$. 
The fourth model, $M_3$, is a constant jerk parametrization, $j(z)=j_0$ 
(examined for the first time by \cite{riess04}).
As one may show, the cosmic jerk is related with the
deceleration parameter by the differential equation
\begin{equation}
  j = - \left[ q + 2q^{2} + (1 + z)\frac{dq}{dz} \right] \,.
  \label{jerk_relation}
\end{equation}
It should be stressed that kinematic models with constant jerk parameter 
are also very attractive because the historically important dynamical EdS 
cosmology and, with more generality, the flat $\Lambda$CDM scenario are 
particular cases for which $j(z)=j_0=-1$.

On the other hand, it is widely known that if one wishes to
describe the recent cosmic expansion, then the current values 
of the parameters given by
(\ref{hubble}), (\ref{dec}) and (\ref{jerk}) lead to
the late time cosmic expansion \cite{Weinb72,Visser04} 
\be{a_exp}
a(t)= 1+H_0(t-t_0) - \frac{1}{2}q_0 H_0^2(t-t_0)^2 -
\frac{1}{3!}j_0 H_0^3(t-t_0)^3 + {\cal O}[(t-t_0)^4] ,
\ee 
from which the luminosity distance can be expanded, yielding 
an extended version of the Hubble law \cite{1998PThPh.100.1077C}
\be{dL_exp}
d_L(z)=\frac{c}{H_0}\left[z +
\frac{1}{2}(1-q_0)z^2-\frac{1}{6}(1-q_0-3q_0^2-j_0)z^3\right] 
+{\cal O}(z^4),
\ee
where the highest order term depends on fourth order
and higher derivatives of the scale factor. 
A shortcoming of this series is that for supernovae at $z>1$ the 
${\cal O}(z^4)$ terms may be large in principle, 
{\it i.e.}, there is a convergence problem \cite{2007gr.qc.....3122C}.
Nevertheless, most of the known supernovae are at $z<1$ and 
a truncation of (\ref{dL_exp}) always can be seen as a polynomial fit.
So we define our last model, $M_4$, as the third order truncation of 
(\ref{dL_exp}), which also has as free parameters $q_0$ and $j_0$. 
However, differently from $M_3$, this model has variable jerk parameter 
(see \ref{jerk_M4}).
 
In the Appendix, one may find the basic analytical expressions for all 
models investigated in the present work.

\section{Sample and Statistical Analysis}

In the statistical analysis below we consider the most complete
supernovae data set currently available, namely, the Union supernova
sample as compiled by Kowalski et al. \cite{kowalski2008}. 
The Union SNIa compilation is a new data set at low and intermediate
nearby-Hubble-flow redshifts whose analysis procedures permit to
work with several heterogeneous supernova samples. It includes 13
independent sets, and, after selection cuts, the robust compilation
obtained is composed by 307 SNIa events distributed over the
redshift interval $0.015 \leq z \leq 1.62$.

In figure 1 we show the behaviour and dispersion of the data 
in the reduced Hubble-Sandage diagram.
The curves correspond to the models considered in this work, 
as indicated in the legend. 
For the sake of comparison, we also show the 
present cosmic concordance model.

\begin{figure}
\centering
\includegraphics[scale=0.56]{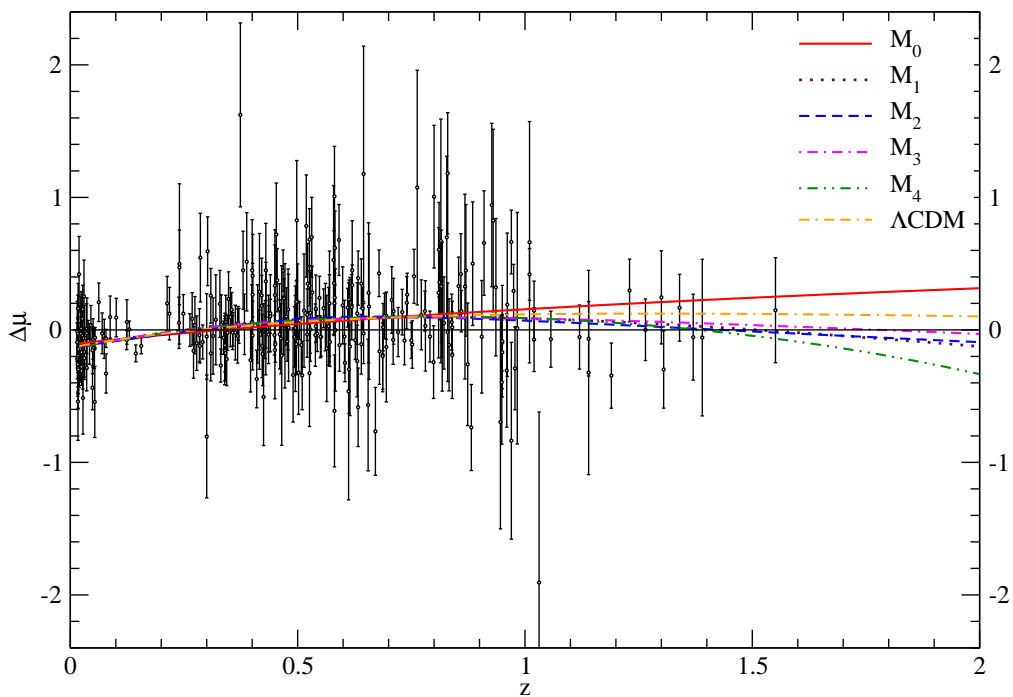}
\fl
\caption{Kinematic model predictions for SNe Ia data. 
Residual magnitude versus redshift is displayed for 307 SNe type Ia from 
SCP union compilation. The predictions of five kinematic models ($M_0-M_4$) 
are displayed relative to an eternally coasting model, for which 
$H(z)=H_0(1+z)$ and $q(z)\equiv 0$. 
For comparison, the cosmic concordance model has also been included.
Note that for $z > 1.5$, the model with $j$ is a free parameter constant 
($M_3$) is the closest one to $\Lambda$CDM.} \label{data}
\end{figure}

The distance moduli $\mu$ is defined as the difference between the 
apparent ($m$) and absolute ($M$) magnitudes, so that the observed 
and theoretical values are respectively
\begin{equation}
\mu_{obs,i} = m_{obs,i}-M \;,
\end{equation}
and
\begin{equation}
\mu_{th}(z_i) = m_{th}(z_i)-M=5\log_{10}d_L(z_i,p)+\mu_0 \;,
\end{equation}
where $\mu_0=25-5\log_{10} H_0$, and $d_L$ is the luminosity distance
\begin{equation}
d_L(z,p)= (1+z)\int_0^z\frac{du}{H(u)} \; , \label{lumin_dist}
\end{equation}
which carries the model parameter dependencies represented by $p$.

The likelihood analysis is based on the calculation of
\begin{eqnarray}
  \chi^2(p,\mu_0) 
  & \equiv & \sum_{SNIa}\frac{\left[
      \mu_{obs,i}-\mu_{th}(z_i)\right]^2} {\sigma_i^2} \\
  & = &\sum_{SNIa}\frac{\left[ \mu_{obs,i} - 5 \log_{10} d_L(z_i,p) - \mu_0
      \right]^2} {\sigma_i^2} \; . 
  \label{chi2}
\end{eqnarray}
We analytically marginalize over the nuisance parameter $\mu_0$
\cite{rapp07},
\begin{equation}
\tilde{\chi}^2(p) = -2 \ln \int_{-\infty}^{+\infty}\exp \left[
-\frac{1}{2} \chi^2(p,\mu_0) \right] d\mu_0 \; ,
\label{chi2_marginalization}
\end{equation}
to obtain
\begin{equation}
\tilde{\chi}^2 =  a - \frac{b^2}{c} + \ln \left(
\frac{c}{2\pi}\right) , \label{chi2_marginalized}
\end{equation}
where
\begin{equation}
a=\sum_{SNIa} \frac {\left[ 5 \log_{10}
d_l(z_i,p)-\mu_{obs,i}\right]^2}{\sigma_i^2} ,
\end{equation}
\begin{equation}
b=\sum_{SNIa} \frac {5 \log_{10} d_l(z_i,p)-\mu_{obs,i}}{\sigma_i^2} ,
\end{equation}
\begin{equation}
c=\sum_{SNIa} \frac {1}{\sigma_i^2} \; .
\end{equation}

The nuisance parameter value that minimizes (\ref{chi2}) is
$\mu_0=b/c$. The expression $\chi^2(p,b/c)=a-(b^2/c)$ is sometimes
used instead of (\ref{chi2_marginalized}) to perform the
likelihood analysis. Both are equivalent if the prior for $\mu_0$ is
flat, as is implied in (\ref{chi2_marginalization}), and the
errors $\sigma_i$ are model independent, what also is the case here.
For the SNIa sample used we find 
$\tilde{\chi}^2(p)-\chi^2(p,b/c)=\ln(c/2\pi)\approx 7.2$.

To determine the best fit parameters for each model, we minimize
$\tilde{\chi}^2(p)$, what is equivalent to maximizing the likelihood
\begin{equation}
{\cal{L}}(p) \propto e^{-\tilde{\chi}^2(p)/2} .
\end{equation}

The Bayesian evidence can then be calculated as
\begin{equation}
E \equiv \int {\cal{L}}(p) P(p) dp
= \frac{1}{V_P}\int_{V_P} {\cal{L}}(p) dp ,
\label{Bevidence}
\end{equation}
where $P(p)$ is the prior probability distribution for the
parameters, which we adopt to be flat,
and $V_{P}$ is the volume in the parameter space defined
by the prior intervals.
We chose conservative prior intervals (see table \ref{table}) 
based on physical considerations and ``prior'' information,
{\it i.e.} previous results obtained with older and smaller SNIa samples
\cite{shapiro,EM,rapp07,2007gr.qc.....3122C}. 
The prior boundaries are chosen to be large enough so most of the 
likelihood is retained in the integration (\ref{Bevidence}), but not 
too large to do not excessively penalize the Bayesian evidence 
through $V_P$.
For all the models considered here, except for $M_2$, as will be 
discussed in the next section, the $3\sigma$ boundaries in the 
likelihood are well inside the prior volume, so the Bayesian evidence 
decreases linearly with $V_P$.

We are able to compare models by calculating the Bayes factor
between any two models $M_i$ and $M_j$, which we define as
\begin{equation}
B_{ij}=\frac{E(M_j)}{E(M_i)}.
\label{BayesFactor}
\end{equation}
Some authors \cite{jeffreys,trotta} offer qualitative 
interpretations of the Bayes factor value (Jeffreys scale) that say how 
one model is favoured over the other, given the data and priors. 
Note that under our convention of the Bayes factor, if 
$E(M_j) > E(M_i)$ than $\ln B_{ij}$ is positive.

\section{Results and Discussion}

\begin{figure}
\centering
\includegraphics[scale=0.4]{figure2a.eps}
\hfill
\includegraphics[scale=0.8]{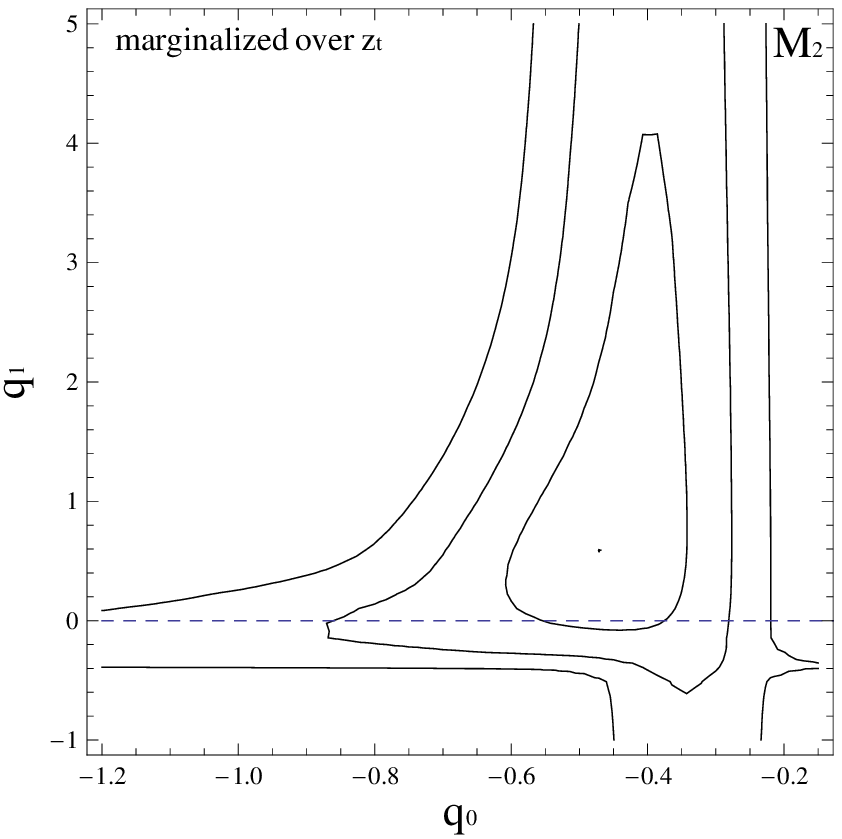}
\\
\includegraphics[scale=0.8]{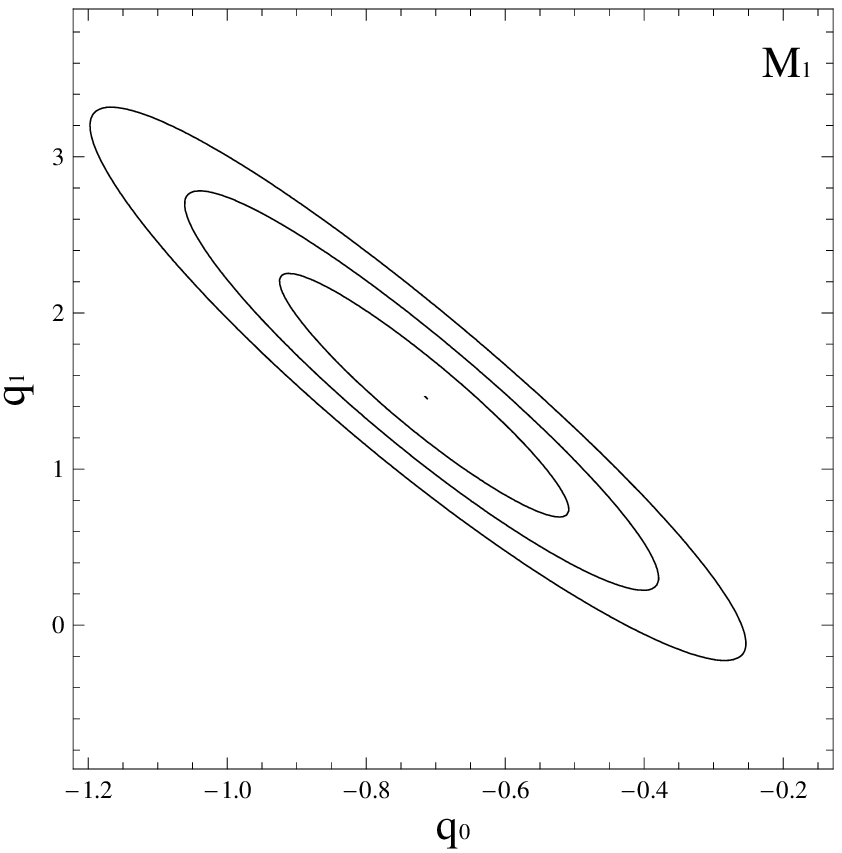}
\hfill
\includegraphics[scale=0.81]{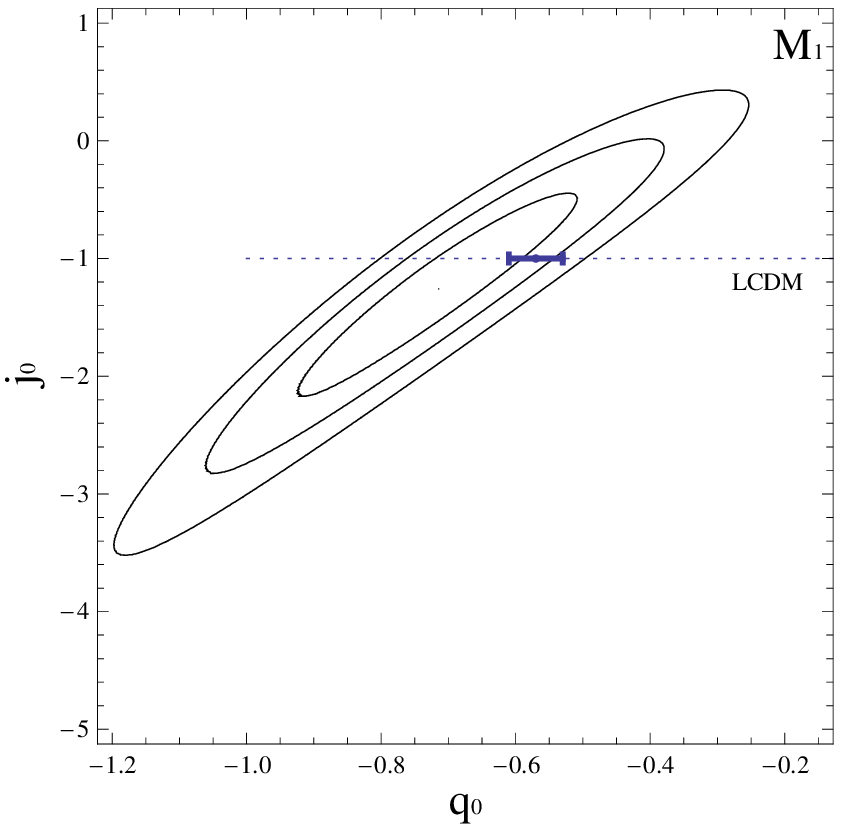}
\\
\includegraphics[scale=0.81]{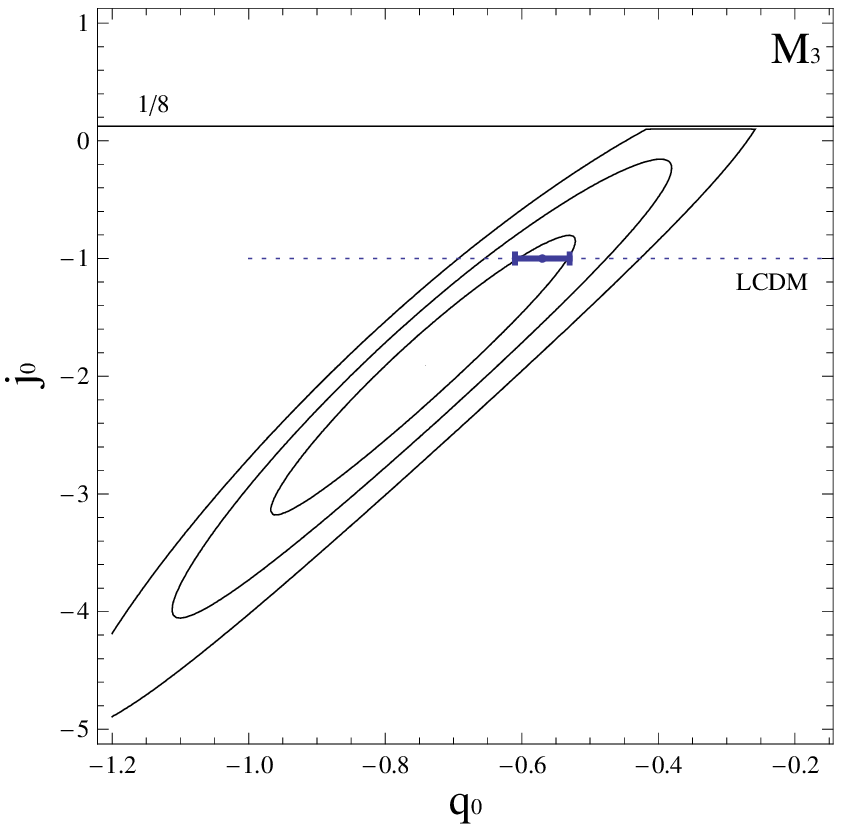}
\hfill
\includegraphics[scale=0.81]{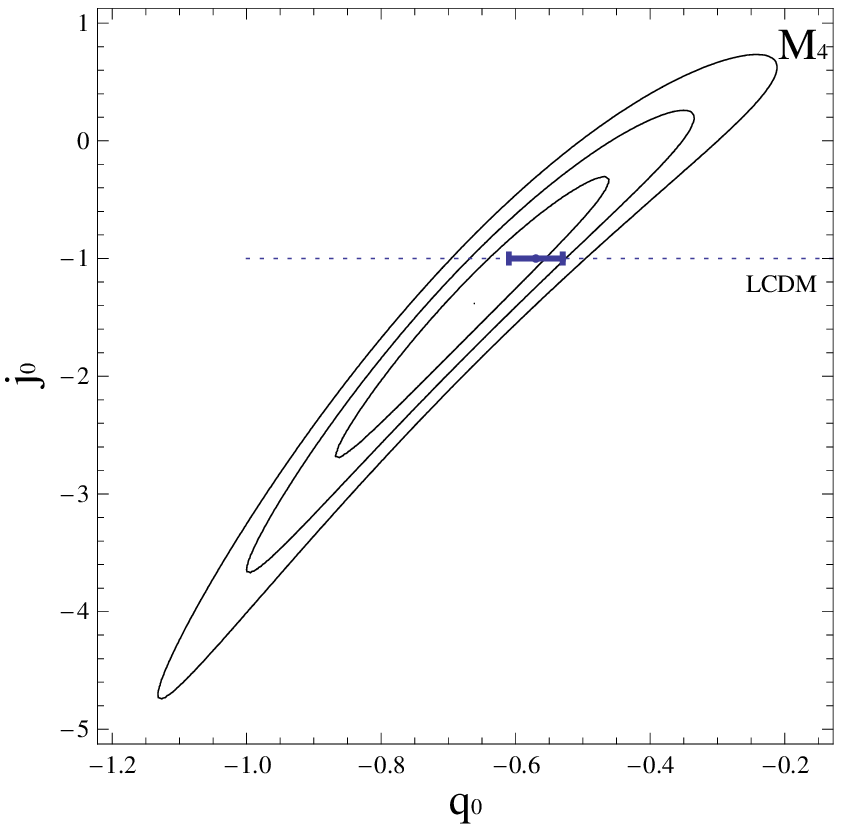}
\caption{
Likelihood results for all studied models (labels on the top of each panel).  
The top left panel is for models with one free parameter 
($\Lambda$CDM and $M_0$). 
The remaining panels show the likelihood contours in the 
parameter space of each model, delimiting the $1\sigma$, 
$2\sigma$ and $3\sigma$ confidence regions. 
Model $M_1$ is also shown in the ($q_0$,$j_0$) parameter space.
The dotted line in three of the panels depicts the image of 
$\Lambda$CDM models in the ($q_0$,$j_0$) parameter space representation 
and the error bar corresponds to the $1\sigma$ region for the best 
$\Lambda$CDM fit. 
The solid straight lines on the panels for $\Lambda$CDM and $M_3$ depict 
the allowed regions in parameter space for these models.}
\label{likel1}
\end{figure}

In figure \ref{likel1} we show the likelihood results for the 
kinematic models considered and also for the $\Lambda$CDM model 
for comparison. 
The horizontal axis depicts the deceleration parameter 
today in the same scale for all models to facilitate the comparison 
among models.
At the panels for $\Lambda$CDM and $M_0$ the full likelihood is plotted 
as a function of $q_0$ --- we use that $q_0=\frac{3}{2}\Omega_m -1$ for 
$\Lambda$CDM, see (\ref{lcdm_q}) --- since these models have only one 
free parameter. 
For the other models it is presented the confidence contours in 
two dimensional parameter spaces (we marginalize over $z_t$ for 
model $M_2$ that has three degrees of freedom). 
The equations for $M_3$ --- see (\ref{M3alpha}) in the Appendix --- 
indicate that the maximum physical value for $j_0$ in this model is $1/8$, 
so we limit the graph to this region.

Table \ref{table} contains the maximum likelihood values and 1$\sigma$ 
projected errors for the deceleration and jerk parameter today and for the 
transition redshift. 
Note that some of these parameters are the best fit values for free 
parameters of the models and others are derived quantities from them. 
See the Appendix for the expressions.
Table \ref{table} also presents the parameter space volume of the priors 
for each model; the goodness of fit as quantified by 
$\tilde{\chi}^2_{red} \equiv \tilde{\chi}^2_{min}/(N-n_p)$, 
where $N$ is the number of data points and $n_p$ is the number of free 
parameters in each model; and the Bayes factor in relation to model $M_0$.

\begin{table}
\begin{center}
\begin{tabular}{lcccrcl}
\hline Model & $q_0$ & $j_0$ & $z_t$ & $V_{P}$ & $\tilde{\chi}^2_{red}$ & 
$\ln B_{0i}$ \\
\hline ${\Lambda CDM}$ & $-0.57\pm0.04$ & $-1$ & $0.71^{+0.08}_{-0.07}$ & 1 & 1.043 & 3.6 \\
$M_0$ : $q=q_0$ & $-0.34\pm0.05$ & {\small 0.11$\pm$0.02}  & {\tiny no transition} & 2 & 1.065& 0 \\
$M_1$ : $q=q_0+q_1 z$ & $-0.71\pm0.21$ & $-1.3^{+0.8}_{-0.9}$ & $0.49^{+0.27}_{-0.09}$ & 10 & 1.040& 3.5 \\
$M_2$ : $q= \left\{
\begin{array}{ll}
q_0,\, z\leq z_t\\
q_1,\, z>z_t\\
\end{array}
\right.$
& $-0.49^{+0.13}_{-0.26}$ & $0.01^{+0.09}_{-0.38}$ & $0.46^{+0.40}_{-0.28}$ & 18 & 1.043& 2.1\\
$M_3$ : $j=j_0$ & $-0.74\pm0.22$ & $-1.9^{+1.1}_{-1.3}$ & $0.48^{+0.36}_{-0.11}$ & 10 & 1.041& 3.3\\
$M_4$ : $d_L(z)$ expansion & $-0.66\pm0.20$ & $-1.4^{+1.1}_{-1.3}$ & $0.52^{+0.21}_{-0.08}$ & 12 & 1.040& 3.2 \\
\hline
\end{tabular}
\end{center}
\caption{Models, maximum likelihood estimates of the universal kinematic 
parameters ($q_0$, $j_0$, $z_t$) and $1\sigma$ projected errors, prior volume, 
goodness of fit and Bayes factor in relation to $M_0$. 
All models are in a flat universe, $V_{P}$ is the prior volume; 
priors are $q_0\in[-2,0]$ for all kinematic models, 
$q_1\in[-1,4]$ for $M_1$, $q_1\in[-1,5]$ and $z_t\in[0,1.5]$ for $M_2$, 
$j_0\in[-5,1/8]$ for $M_3$, $j_0\in[-5,1]$ for $M_4$, and 
$\Omega_m \in[0,1]$ for $\Lambda$CDM. 
The relations between the particular model parameters and the universal 
kinematic parameters are given in the Appendix.} 
\label{table}
\end{table}

The deceleration today is significantly negative in all models, 
but its exact value and uncertainty are model-dependent. 
The likelihoods for models with just one degree of freedom 
($M_0$ and $\Lambda$CDM) are very peaked in $q_0$ 
(see top left panel of figure \ref{likel1}), yielding 
a more precise determination of this parameter, and consequentially also of 
$j_0$, than in the other models (see table \ref{table}).
For the models of constant acceleration today ($M_0$ and $M_2$) the 
current values of the cosmic jerk are in disagreement with the values 
found for the remaining models, which are compatible among themselves and 
with the $\Lambda$CDM prediction.
The values of the transition redshift determined in all kinematic models 
are close to $z_t\sim0.5$, and are compatible in $1\sigma$ with the higher 
value obtained in the $\Lambda$CDM model.

Our likelihood contours for model $M_1$ at figure \ref{likel1} are 
qualitatively similar, but with tighter constraints, to what was 
obtained in previous works \cite{riess04,EM}. 
We also represent the $M_1$ model in the parameter space of the 
deceleration and jerk today, making use of (\ref{M1_jerk}).

The panel for $M_2$ shows a qualitatively similar plot to the last 
panel of figure 2 of Shapiro \& Turner \cite{shapiro}.
The $z_t$ marginalized likelihood contours for $M_2$ at $2\sigma$ and 
$3\sigma$ confidence levels show that the data do not constrain strongly 
the deceleration parameter for redshifts above the transition.
The parameter $q_1$ can assume very large positive values and also 
negative values. 
In this last case, represented by the points below the dashed line on 
the $M_2$ panel of figure \ref{likel1}, $z_t$ is a transition redshift 
between two accelerated phases. 
Points above the dashed line represent situations in which there is a 
transition from a decelerated phase at high $z$ to an accelerated phase 
at low $z$.
We also tested a larger upper prior boundary for $q_1$, but even for $q_1=10$ 
the $2\sigma$ likelihood contour is open (in this case $\ln B_{02}=1.6$).

Models $M_3$ and $M_4$ (bottom panels at figure \ref{likel1}), 
and $M_1$ as well, have roughly similar likelihood contours in the 
($q_0$,$j_0$) plane due to the relation (\ref{jerk_relation}), 
implying similar determinations of $q_0$ and $j_0$, similar goodness of fit 
and Bayesian evidences.
These likelihood contours also show that the best fit $\Lambda$CDM model, 
represented by a $1\sigma$ error bar over the dotted line with 
$j_0=-1$, is compatible at $1\sigma$ confidence level with $M_1$, $M_3$ 
and $M_4$. 
Similar result was obtained by Rapetti et al. \cite{rapp07} using 
a joint analysis of the Gold \cite{riess04} and SNLS \cite{snls} SNIa samples 
and X-ray cluster gas mass fraction measurements. 
When considering separately the SNIa sets, those authors found 
$j_0=-2.8^{+1.1}_{-1.2}$ for the Gold set and 
$j_0=-1.3^{+1.2}_{-1.4}$ for the SNLS set.
We also examined the case where just supernovae with $z<1$ (289 events) 
were fitted by model $M_4$ and the results do not differ considerably 
from the ones for the full sample (307 events). In fact, the goodness 
of fit is worst for the subsample with $z<1$, $\tilde{\chi}^2_{red}=1.057$.

All models, except $M_0$, have similar goodness of fit, as quantified by 
$\tilde{\chi}^2_{red}$. In fact, if we observe the best fit curves at 
figure \ref{data}, it is very difficult to judge which one best describes 
the data. The Bayes factor segregates $M_0$ from the other models, but only 
weakly disfavours $M_2$ over the remaining models, mainly because of 
the penalizing effect of the larger prior volume of $M_2$ 
(due to it having three free parameters) in relation to the other models.
The basic result is that the Bayes factor is unable to significantly 
distinguish models $M_1$, $M_3$, $M_4$ and $\Lambda$CDM.
This is a somewhat different result from what was found by 
Elgar{\o}y \& Multam{\"a}ki \cite{EM} when examining separately 
the Gold \cite{riess04} and SNLS \cite{snls} SNIa sets. 
There the authors obtained a clear ranking of the models based on the
Bayes factor (see their table 5), even though these rankings were 
different for the two SNIa samples.

It is worth to note that the behaviour for the evolution of the 
deceleration parameter is very distinct among the models considered here. 
The simplest kinematic models, $M_0$, $M_1$ and $M_2$, 
have self-evident $q(z)$, but $M_3$ has a more subtle, and interesting 
acceleration history. 
For this model  $q(z)$ --- given by (\ref{M3_q}) --- has a qualitatively 
similar behaviour to that of $\Lambda$CDM (\ref{lcdm_q}), namely, 
asymptotic constant values in the past and future 
with a smooth transition around $z_t$.
For $\Lambda$CDM the asymptotic limits are fixed, 
$q(z\rightarrow \infty)=0.5$ and 
$q(z\rightarrow -1)=-1$, but for $M_3$ they will depend on $q_0$ and $j_0$. 
In the specific case of our best fit values for $M_3$, 
$q(z\rightarrow \infty)\sim 0.76$ and $q(z\rightarrow -1)\sim -1.3$.
There is no deceleration in the future in this model, and therefore, 
no new phase transition.
For $M_4$ the extrapolation of $q(z)$ is not expected to be valid far 
from the redshift range for which it was adjusted, and, in fact, 
(\ref{M4_q}) presents nonphysical behaviour in the extremes 
past and future.
None of our models are able to hint at any slowing down of the cosmic 
acceleration today or in the future, such as suggested recently 
\cite{shafieloo} in the context of an extended SNIa sample.

\section{Summary and Conclusions}

Bellow we summarize our main accomplishments and conclusions.

1. We perform, for the first time, a kinematic analysis of the 
307 SNIa compiled in the Union set \cite{kowalski2008}. 
This approach, also called Friedmanless, allows us to analyse 
the cosmic expansion just having to assume the homogeneity and 
isotropy of the universe, and not having to make any assumption 
about the underlying gravitational theory and energy components 
of the universe.

2. We employ several kinematic models that were used by different 
groups with various SNIa samples before. Using a unified framework 
and a single data set, we are able to compare the kinematic models 
directly. We calculate for each model the goodness of fit 
(as measured by $\chi^2_{red}$) and the Bayes factor.
Even very simplistic kinematic models can give an equivalent description 
of the cosmic expansion to the one provided by the currently favoured 
concordance $\Lambda$CDM model. 
More to the point, current data is not powerful enough to clearly 
discriminate among some of these simple models. 
This is a distinct conclusion from what Elgar{\o}y \& Multam{\"a}ki \cite{EM} 
obtained using separately the SNLS and Gold samples for a particular class 
of the models studied here, and who found conflicting results from the 
two samples.
Nevertheless, some kinematic models ($M_1$, $M_3$ and $M_4$) were shown 
to be superior, or more realistic, than kinematic models with constant 
deceleration ($M_0$ and $M_2$).

3. We give, for all kinematic models studied, the expressions and 
estimates for a minimal set of parameters that characterizes the recent 
history of the cosmic expansion:
$q_0$ (deceleration today), $j_0$ (cosmic jerk today) and 
$z_t$ (transition redshift from a decelerated to an accelerated phase).

4. Independently of models, the universe is in a phase of
accelerated expansion, however the value of the deceleration
parameter today is model-dependent. For the most realistic kinematic models 
($M_1$, $M_3$ and $M_4$), $q_0$ is in the $1\sigma$ range $[-0.96,-0.46]$. 

5. There is evidence that the deceleration parameter was higher and
positive in the past, implying a transition from a decelerated phase
to an accelerated one. The transition redshift between these two
phases is found to be around 0.5 in all kinematic models, being
in the $1\sigma$ range $[0.36,0.84]$ for the the most realistic kinematic 
models. That is compatible in $1\sigma$ with the higher value predicted 
by $\Lambda$CDM.

6. The value of the cosmic jerk today can be used as a
measure of a possible deviation from a $\Lambda$CDM model, which 
exactly predicts $j_0=-1$ by definition.
For the models with constant deceleration today ($M_0$ and $M_2$) 
the value for $j_0$ is significantly higher than for $\Lambda$CDM, 
but these models no longer have a strong cosmological appeal.
The remaining kinematic models ($M_1$, $M_3$ and $M_4$) have a 
slightly preference for $j_0<-1$, being in the $1\sigma$ range $[-3.2,-0.3]$.
But at the current confidence level yielded by the data there is no 
significant departure from the $\Lambda$CDM prediction.
We note that the constant jerk model is an interesting parameterization 
of the cosmic expansion because, among other reasons, it contains popular 
models as $\Lambda$CDM and EdS. 

Finally, taking into account the discussion in the previous section, 
about the behaviour of the deceleration parameter in the context of 
jerk models, it could be interesting to extend the present work by 
allowing additional contributions from a snap parameter 
(dependent on the fourth order derivative of the scale factor).
Either in a simplified model, $s(z)=constant$, as well as in a 
expansion of the luminosity distance. 
Hopefully, this may help us to have some indication about a possible 
dynamic transition in the future, with the universe entering in a 
new decelerating phase \cite{Lima2}.  
Some work along these lines will presented in a forthcoming communication.

\section*{Acknowledgements}
ACCG is supported by  FAPESP under grant 07/54915-9.
JVC is supported by FAPESP under grant 05/02809-5 and partially also 
by CNPq under grant 477190/2008-1.
JASL is partially supported by CNPQ and FAPESP under grants
304792/2003-9 and 04/13668-0, respectively.

\appendix
\section{Kinematic Expressions}

In this appendix we show the analytical expressions for the basic
quantities in all models investigated in the present work.

\vspace{0.5cm}

$\bf M_0$
\begin{equation}
 H(z) = H_0(1+z)^{1+q_0}
\end{equation}
\begin{equation}
 q(z) = q_0
\end{equation}
\begin{equation}
 j(z) = -(q_0+2q_0^2)
\end{equation}

$\bf M_1$
\begin{equation}
 H(z)=H_0(1+z)^{1+q_0-q_1 }e^{q_1 z}
\end{equation}
\begin{equation}
 q(z) = q_0+q_1 z
\end{equation}
\begin{equation}
 j(z)= -\left[ q(z) + q^2(z) + (1+z) q_1 \right] \label{M1_jerk}
\end{equation}
\begin{equation}
 z_t=-q_0/q_1
\end{equation}

$\bf M_2$
\begin{equation}
 H(z)=\left\{
 \begin{array}{ll}
   H_0(1+z)^{1+q_0}, \, z\leq z_t\\
   H_0(1+z_t)^{q_0-q_1}(1+z)^{1+q_1}, \, z>z_t\\
 \end{array}
 \right.
\end{equation}
\begin{equation}
 q(z)= \left\{
 \begin{array}{ll}
   q_0,\, z\leq z_t\\
   q_1,\, z>z_t\\
 \end{array}
 \right.
\end{equation}
\begin{equation}
 j(z) = \left\{
 \begin{array}{ll}
   -(q_0+2q_0^2),\, z\leq z_t\\
   -(q_1+2q_1^2),\, z>z_t\\
 \end{array}
 \right.
\end{equation}

$\bf M_3$
\begin{equation}
 H(z)= H_0 [ c_1(1+z)^{\alpha_1} + c_2(1+z)^{\alpha_2} ]^\half 
\label{M3huble}
\end{equation}
\begin{equation}
q(z)= \frac{c_1(1+z)^{\alpha_1}(\frac{\alpha_1}2-1) + c_2(1+z)^{\alpha_2}(\frac{\alpha_2}2-1)}{c_1(1+z)^{\alpha_1}+c_2(1+z)^{\alpha_2}}
\label{M3_q}
\end{equation}
\begin{equation}
 j(z)=j_0
\end{equation}
\begin{equation}
 z_t= \left[ - \frac{c_2}{c_1}\frac{\alpha_2-2}{\alpha_1-2}\right]^
 \frac{1}{\alpha_1-\alpha_2}-1
 \label{M3zt}
\end{equation}
where
\begin{equation}
 \alpha_{1,2}=\frac{3}{2}\pm\sqrt{\frac{9}{4}-2(1+j_0)}
 \label{M3alpha}
\end{equation}
\begin{equation}
c_1=\frac{2(1+q_0)-\alpha_2}{\alpha_1-\alpha_2} 
\;\;\;\;\;\;\; {\rm and} \;\;\;\;\;\;\;  c_2=1-c_1
\end{equation}
From (\ref{M3alpha}) we see that $j_0<\frac{1}8$.

{$\bf M_4$} -- defined by the expanded luminosity distance (\ref{dL_exp}), 
$d_L(z)=\frac{c}{H_0} \left( z+ Az^2+ Bz^3 \right)$, where
$A=(1-q_0)/2$ and $B=-(1-q_0-3q_0^2-j_0)/6$.
\begin{equation}
  H(z)= H_0 \left[ \frac{(1 + z)^2}
    {1+2Az+(A+3B)z^2+2Bz^3} \right]
\end{equation}
\begin{equation}
  q(z)= \frac{1-2A-2(A+3B)z-(A+9B)z^2-2Bz^3}{1+2Az+(A+3B)z^2+2Bz^3}
  \label{M4_q}
\end{equation}
\begin{equation}
  j(z)= -\left[ q + 2q^2 + (1+z) q^{\prime} \right]
  \label{jerk_M4}
\end{equation}
\begin{equation}
  z_t: {\rm the \; real \; root \; of \;} \;
  1-2A-2(A+3B)z_t-(A+9B)z_t^2-2Bz_t^3=0
\end{equation}

{\bf $\bf \Lambda$CDM}, $\Omega_m+\Omega_\Lambda=1$
\begin{equation}
 H(z)= H_0 \left[ \Omega_m (1+z)^3 + (1-\Omega_m)\right]^\half
\end{equation}
\begin{equation}
 q(z)= \left[(1+z)^3-2(1/\Omega_m-1)\right]
 / \left[2(1+z)^3+2(1/\Omega_m-1)\right]
\label{lcdm_q}
\end{equation}
\begin{equation}
 j(z)= -1
\end{equation}
\begin{equation}
 z_t= [2(\Omega_m^{-1}-1)]^{\frac{1}3}-1
\end{equation}
Note that the expressions for $\Lambda$CDM can be easily obtained from 
(\ref{M3huble}-\ref{M3zt}), putting $j_0=-1$ into (\ref{M3alpha}).
If we do similarly for $M_0$, putting $j_0=-(q_0+2q_0^2)$ into 
(\ref{M3alpha}), we recover the right expressions for $H(z)$ and $q(z)$, 
and, curiously, $z_t=-1$. 
This just illustrates that all models with a particular constant jerk 
are particular cases of $M_3$.



\section*{References}
\begin{thebibliography}{X}
\bibitem{1998AJ....116.1009R} Riess A G {\it et al.} 1998 Astron.\ J.\ {\bf 116} 1009 [astro-ph/9805201]
\bibitem{perlmutter} Perlmutter S {\it et al.} 1999 Astrophys. J. {\bf 517} 565 [astro-ph/98121330
\bibitem{riess04} Riess A G ~{\it et al.} 2004 Astrophys.\ J.\  {\bf 607} 665 [astro-ph/0402512] 
\bibitem{snls} Astier P {\it et al.} 2006 Astron. \& Astrophys. {\bf 447} 31 [astro-ph/0510447] 
\bibitem{rev1} Peebles P J E and Ratra B 2003 Rev.~Mod.~Phys. {\bf 75} 559 [astro-ph/0207347]
\bibitem{rev2} Padmanabhan T 2003 Phys.~Rept. {\bf 380} 235 [hep-th/0212290]
\bibitem{rev3} Lima J~A~S 2004 Braz.~Journ.~Phys. {\bf 34} 194 
[astro-ph/0402109]
\bibitem{rev4} Copeland E~J M  and Tsujikawa S 2006
Int.~J.~Mod.~Phys. {\bf D15} 1753 [hep-th/0603057]
\bibitem{rev5} Frieman  J~A, Turner M~S and ~Huterer D 2008
Ann.~Rev.~Astron. \& Astrophys. {\bf 46} 385 
\bibitem{lima96} Lima J A S, Germano A~S~M. and Abramo L~R~W 1996 
Phys.~Rev.~D {\bf 53} 4287 
\bibitem{lima08} Lima J~A~S, Silva F~E and Santos R~C~ 2008
Class.~Quant.~Grav. {\bf 25} 205006
\bibitem{gary08} Steigman G, Santos R C and Lima J A S 2009 JCAP {\bf 6} 33 
\bibitem{FR} Barrow J D and Cotsakis S 1988 Phys. Lett. B {\bf 214} 515
\bibitem{FR1} Li B and Barrow J D 2007 Phys. Rev. D {\bf 75} 084010 [gr-qc/070111]
\bibitem{FR2} Amarzguioui M, Elgar{\o}y {\O}, Mota D F and Multamaki T 2006 [astro-ph/0510519]
Astron. Astrophys. {\bf 454} 707
\bibitem{FR3} Fairbairn M and Rydbeck S 2007 JCAP {\bf 0712} 005[astro-ph/0701900] 
\bibitem{FR4} Carvalho F C, Santos E M, Alcaniz J S and Santos J 2008 JCAP {\bf 9} 8 [arXiv:0804.2878]
\bibitem{Komat08} Komatsu E {\it{et al.}} (WMAP Collaboration) 2009 Astrophys. J. Suppl. {\bf 180} 330 [arXiv:0803.0547] 
\bibitem{TurRie02} Turner  M S and  Riess A G 2002 Astrophys. J. {\bf{569}} 18 [astro-ph/0106051]
\bibitem{Visser04} Visser M 2004 Class. Quant. Grav. {\bf 21} 2603 [gr-qc/0309109]
\bibitem{Visser05} Visser M 2005 General Relativity and Gravitation {\bf 37} 1541 [gr-qc/0411131] 
\bibitem{shapiro} Shapiro C and Turner M S 2006 Astrophys. J. {\bf 649} 563 [astro-ph/0512586] 
\bibitem{blandford05} Blandford R D, Amin M, Baltz E A, Mandel K and 
Marshall P J 2005 Observing Dark Energy, 339, 27 [astro-ph/0408279]
\bibitem{EM1} Elgar{\o}y {\O} and Multam\"{a}ki T 2005
Mon.\ Not.\ Roy.\ Astron.\ Soc.\  {\bf 356} 475 [astro-ph/0404402] 
\bibitem{EM} Elgar{\o}y {\O} and Multam{\"a}ki T\ 2006 JCAP {\bf 9} 2 [astro-ph/0603053]
\bibitem{Virey05} Virey J-M  {\it{et al.}} 2005 Phys. Rev. D {\bf 72} R061302
\bibitem{rapp07} Rapetti D, Allen S W, Amin M A. and Blandford R~D 2007 
MNRAS {\bf 375} 1510 [astro-ph/0605683]
\bibitem{daly08} Daly R~A {\it et al.} 2008 Astrophys.\ J.\  {\bf 677} 1 
\bibitem{CL08} Cunha J V and Lima J A S 2008 
Mon. Not. R. Astron. Soc. {\bf{390}} 210 [arXiv:0805.1261]
\bibitem{Cunha09} Cunha J V 2009 Phys. Rev. D {\bf 79} 047301 [arXiv:0811.2379]
\bibitem{kowalski2008} Kowalski M {\it et al.} 2008  Astrophys. J. {\bf 686} 749 [arXiv:0804.4142]
\bibitem{Weinb72} Weinberg S 1972 {\it{Cosmology and Gravitation}}(John Wiley Sons, New York)
\bibitem{1998PThPh.100.1077C} Chiba T, Nakamura T 1998, Progress of Theoretical Physics {\bf 100} 1077 
\bibitem{2007gr.qc.....3122C} Catto{\"e}n C, Visser M\ 2007, arXiv:gr-qc/0703122
\bibitem{jeffreys} Jeffreys H 1961 {\em Theory of Probability} 
(Oxford: Clarendon Press)
\bibitem{trotta} Trotta R 2007 Mon.\ Not.\ Roy.\ Astron.\ Soc. {\bf 378} 72 [astro-ph/0504022]
\bibitem{shafieloo} Shafieloo A, Sahni V, and Starobinsky A~A 2009 arXiv:0903.5141
\bibitem{Lima2} Carvalho F C, Alcaniz J S, Lima J A S, Silva R 2006 Phys. Rev. Lett. {\bf 97} 081301 [astro-ph/0608439]
\end{thebibliography}
\end{document}